\def\be{\begin{equation}}
\def\ee{\end{equation}}
\def\ba{\begin{array}}
\def\ea{\end{array}}

\def\Nb{{I\!\! N}}
\def\Rb{{I\!\! R}}

\def\Cb{{\Bbb C}}

\documentstyle[12pt]{article}
\begin{document}
\input amssym.def
\baselineskip=22pt \setcounter{page}{1}
\centerline{{\Large\bf A Note on Entanglement of Formation}}
\medskip
\centerline{{\Large\bf and Generalized Concurrence}\footnote [1]{Supported by the NSF of
China, Grant No. 10271081.}}
\begin{center}
Shao-Ming Fei$^{\dag\ddag}$, Zhi-Xi Wang$^{\dag}$, Hui Zhao$^{\dag}$
\end{center}
\bigskip

\begin{center}
\begin{minipage}{6in}
\parskip=6pt

$^\dag$Department of Mathematics, Capital Normal University,
Beijing 100037, P.R. China

$^\ddag$Institut f{\"u}r Angewandte Mathematik, Universit{\"a}t
Bonn, 53115 Bonn, Germany

\end{minipage}
\end{center}

\vskip 2 true cm
\parindent=18pt
\parskip=6pt
\begin{center}
\begin{minipage}{5in}
\vspace{3ex} \centerline{\large Abstract}
\vspace{4ex}
We discuss a kind of generalized concurrence for a class of high
dimensional quantum pure states
such that the entanglement of formation is a monotonically increasing
convex function of the generalized concurrence. An analytical expression
of the entanglement of formation for a class of high
dimensional quantum mixed states is obtained.

\bigskip
\medskip
\bigskip

PACS numbers: 03.65.Bz, 89.70.+c\vfill

\smallskip
Key words: Entanglement of formation, Generalized concurrence

\end{minipage}
\end{center}

\bigskip

Characterized by the simultaneous occurrence of superposition and
correlation in composite systems, quantum entanglement has in
recent years emerged as the key resource in quantum information
processing \cite{bennet98} and resulted in the explosion of
interest in quantum computing and communication \cite{book}. To
quantify entanglement, a number of entanglement measures has been
proposed for bipartite systems. Among them, the pioneering
contributions of Bennett et al. \cite{bennet96,BBPS} defined
entanglements of formation and distillation on considerations of
convertibility vis-\`a-vis maximally entangled pairs. The relative
entropy of entanglement \cite{vedral02} is based on
distinguishability from the set of separable states. Recently
there are also discussions of entanglement measure on multipartite
case \cite{part}.

The entanglement of formation is defined for arbitrary
dimensional bipartite systems. Due to the
extremizations involved in the calculation,
so far no explicit analytic formulae for entanglement
of formation have been found for systems larger than a pair of
qubits, except for some special symmetric states \cite{th}.

Let ${\cal H}$ be an $N$-dimensional complex Hilbert space with
orthonormal basis $e_i$, $i=1,...,N$. A pure state on ${\cal
H}\otimes{\cal H}$ is generally of the form,
\begin{equation}\label{psi}
\vert\psi\rangle=\sum_{i,j=1}^N a_{ij}e_i\otimes
e_j,~~~~~~a_{ij}\in\Cb
\end{equation}
with normalization $\sum_{i,j=1}^N a_{ij}a_{ij}^\ast=1$.
The entanglement of formation $E$ is defined to be the entropy of
either of the two sub-Hilbert space ${\cal H}\otimes{\cal H}$
\cite{BBPS},
\be\label{epsi} E(|\psi \rangle) = - {\mbox{Tr}\,}
(\rho_1 \log_2 \rho_1) = - {\mbox{Tr}\,} (\rho_2 \log_2 \rho_2)\,,
\ee
where $\rho_1$ (resp. $\rho_2$) is the partial trace of $\bf
|\psi\rangle\langle\psi|$ over the second (resp. first) Hilbert
space of ${\cal H}\otimes{\cal H}$.
Let $A$ denote the matrix with entries given by $a_{ij}$ in
(\ref{psi}). $\rho_1$ can be expressed as $\rho_1=AA^\dag$.
A general density matrix $\rho$ on ${\cal H}\otimes{\cal H}$
has pure-state decompositions of $|\psi_a \rangle$ of the form (\ref{psi})
with probabilities $p_a$,
\be\label{rho}
\rho = \sum_{a=1}^M p_a |\psi_a \rangle
\langle\psi_a|,~~~~\sum_{a=1}^M p_a =1
\ee
for some $M\in\Nb$. The entanglement of formation for the mixed
state $\rho$ is defined as the average entanglement of the pure
states of the decomposition, minimized over all possible
decompositions of $\rho$,
\be\label{erho} E(\rho) = \mbox{min}\,
\sum_{a=1}^M p_a E(|\psi_a \rangle).
\ee

For the case $N=2$, (\ref{epsi}) can be written as
$$
E(|\psi \rangle)|_{N=2} =h(\frac{1+\sqrt{1-C^2}}{2}),
$$
where $h(x) = -x\log_2 x - (1-x)\log_2 (1-x)$, $C$ is called
concurrence, $C(|\psi \rangle)=2|a_{11}a_{22}-a_{12}a_{21}\vert$.
It is easily verified that $E$ is a monotonically increasing
function of $C$, and hence $C$ can be also taken as a kind of
measure of the entanglement. Calculating (\ref{erho}) is reduced
to calculate the corresponding minimum of $C(\rho) = \mbox{min}\,
\sum_{a=1}^M p_a C(|\psi_a \rangle)$, and an analytical expression
of (\ref{erho}) is obtained \cite{HillWootters}.

For $N\geq 3$, there is no explicit analytical expression for the
entanglement of formation in general.
The concurrences discussed in \cite{concu,note} can be only used to judge
whether a pure state is separable (or maximally entangled) or not
\cite{qsep,separ3}, since the entanglement of formation is no longer a
monotonically increasing function of these concurrences.

Nevertheless, for a special class of quantum states,
certain quantities (generalized concurrence) were found to simplify the
calculation of the corresponding entanglement of formation \cite{sjxg}.
Namely, if $AA^\dag$ has only two non-zero eigenvalues $\lambda_1$
and $\lambda_2$, each with degeneracy $m$, then
\be\label{epsinn}
E(|\psi \rangle)=m \left(-x\log_2 x
- (\frac{1}{m}-x)\log_2 (\frac{1}{m}-x)\right),
\ee
where
$$
x = \frac{1}{2}\left(\frac{1}{m}+\sqrt{\frac{1}{m^2}(1-d^2)}\right)
$$
and $d=2m\sqrt{\lambda_1\lambda_2}$
is a kind of generalized concurrence taking values from
$0$ to $1$. One easily shows that $E(|\psi \rangle)$ is a
monotonically increasing function of $d$. Moreover,
$E(d)$ is a convex function,
$$
\frac{\partial^2 E}{\partial d^2}=\frac{\log \frac{1+\sqrt{1-d^2}}
{1-\sqrt{1-d^2}}-2\sqrt{1-d^2}}{(1-d^2)^{3/2}\log 4}>0,~~~~~\forall~ d\in [0,1]\,.
$$
From the monotonicity and convexity the entanglement of formation
for a class of high dimensional mixed states has been calculated
analytically \cite{sjxg,feili}

In the following we generalize the results in \cite{sjxg} to the case
that $AA^\dag$ has $n\geq 3$ different non-zero eigenvalues. We present
the conditions allowing to derive an explicit lower bound of the
entanglement of formation for such kind of
arbitrary dimensional mixed states and calculate the lower bound.

Let $\lambda_1, \lambda_2, \ldots, \lambda_n$, each with
degeneracy $m$, $mn\leq N$, be the non-zero eigenvalues of
$AA^\dag$. $\lambda_i=\lambda_i(u,v)$, $i=1, 2, \ldots, n$, are
differentiable functions of two real variables $u$ and $v$.
We define $D=mn\sqrt{\lambda_1\lambda_2\cdots\lambda_n}$ to be
the generalized concurrence.

{\sf [Lemma]}. If $\lambda_i=\lambda_i(u,v)$, $i=1, 2, \ldots, n$,
satisfy the following conditions: \be\label{cond}
\sum\limits_i\frac{\partial\lambda_i}{\partial
D}log_2{\lambda_i}<0, \ee then $D$ is a measure of entanglement in
the sense that the entanglement of formation of the corresponding
pure state is a monotonically increasing function of $D$.

{\sf [Proof]}\  The normalization condition of $|\psi\rangle$, $Tr
(AA^\dag) =\sum\limits_im\lambda_i=1$, $\lambda_i\in
(0,\frac{1}{m})$, gives rise to \be\label{nm}
\sum\limits_im\frac{\partial\lambda_i}{\partial D}=0\,. \ee The
entanglement of formation of $|\psi \rangle$ is given by
\be\label{enm} E(|\psi\rangle)=-\sum\limits_i m \lambda_i \log_2
\lambda_i \,. \ee From (\ref{nm}) and (\ref{enm}) we have
\be\label{ded} \frac{\partial E}{\partial D}=-\sum\limits_i
mlog_2\lambda_i \frac{\partial\lambda_i}{\partial D}\,,
\ee
which is positive if the condition (\ref{cond}) is satisfied. Therefore
$E(|\psi \rangle)$ is a monotonically increasing function of $D$.
\hfill $\rule{2mm}{2mm}$

The Lemma defines a class of pure states, for which a generalized
concurrence can be still well defined. Here we have supposed that
all $\lambda_i$ are functions of two real variables. This implies that
there is only one independent variable, accounting to the normalization
condition, and the condition (\ref{cond}) could be plausibly satisfied.
The most simple case is that $\lambda_1=u, \lambda_2=v$, each with
degeneracy $m$, the generalized concurrence is given by
$D=2m\sqrt{uv}$, which is just the case discussed in \cite{sjxg},
where $E(|\psi \rangle)$ is not only a monotonically increasing
but also a convex function of $D$, $\frac{\partial^2E}{\partial D^2}
\geq 0$.

As another example we consider non-zero eigenvalues of $AA^\dag$
such that $\lambda_1=u$, $\lambda_2=u+v$, $\lambda_3=u+2v$,
each with degeneracy $m$, $u$ and $v\in\Rb$ taking values $(0,\frac{1}{3m})$.
The generalized concurrence is given by $D=3m\sqrt{u(u+v)(u+2v)}$.
It is straightforward to verify that $E$ is a monotonically increasing function of $D$,
since
$$\sum\limits_i\frac{\partial\lambda_i}{\partial
D}log_2{\lambda_i}=\frac{1}{3mv\sqrt{3m}}(1-9m^2v^2)^{1/2}log_2{\frac{1-3mv}{1+3mv}}<0.$$
Due to the relation
$$\sum\limits_i\frac{1}{\lambda_i}{(\frac{\partial\lambda_i}{\partial
D})}^2 +\frac{\partial^2\lambda_i}{\partial
D^2}ln{\lambda_i}=\frac{1}{27m^3v^3}(6mv+ln{\frac{1-3mv}{1+3mv}})<0,$$
$E$ is also a convex function of $D$.

As $E(|\psi \rangle)$ is a monotonically increasing and
convex function of $D$, instead of calculating $E(\rho)$,
one may calculate the minimum decomposition (in the sense of (\ref{erho})),
$D(\rho) = \mbox{min}\,\sum_{a=1}^M p_a D(|\psi_a \rangle)$,
to simplify the calculations, as long as $\rho$ has all
decompositions on pure states with their eigenvalues of $AA^\dag$
satisfying (\ref{cond}) in Lemma. Nevertheless, like $E(|\psi \rangle)$,
generally the expression of $D(|\psi \rangle)=mn\sqrt{\lambda_1\lambda_2\cdots\lambda_n}$
could be still quite complicated.

In fact $D(|\psi \rangle)$ is an invariant under local unitary transformations.
Associated with a general pure state $|\psi \rangle$ given in (\ref{psi}),
the following quantities are invariants under local unitary transformations \cite{note,qsep}:
$$
I_0=Tr(AA^\dag)=\sum\limits_{i, j=1}^N a_{ij}a_{ij}^*\,,~~~
I_1=Tr[(AA^\dag)^2]=\sum\limits_{i, j, p, q=1}^N a_{ip}a_{iq}^*a_{jq}a_{jp}^*\,.
$$
The generalized concurrence defined in \cite{note}, \be\label{cn}
C_N=\sqrt{\frac{N}{N-1}(I_0^2-I_1)}=\sqrt{\displaystyle
\frac{N}{2(N-1)}\sum_{i,j,p,q=1}^N \vert
a_{ip}a_{jq}-a_{iq}a_{jp}\vert^2}\,, \ee where $C_2=C$, does have
a very simple form. However, the entanglement of formation is
generally not a monotonically increasing function of $C_N$, and
$C_N$ can be only used to judge wether a state is separable or
maximally entangled for the case $N\geq 3$. If the generalized
concurrence $D=mn\sqrt{\lambda_1\lambda_2\cdots\lambda_n}$
satisfies the Lemma and can be further expressed as
$D=\frac{mn}{\sqrt{2}}\sqrt{{I_0^2-I_1}}$, the calculation of the
corresponding entanglement of formation would be greatly
simplified.

Let $\Psi$ denote the set of all pure states of the form
(\ref{psi}) such that i) the Lemma is satisfied; ii) the
entanglement of formation is a convex function of $D$, i.e.,
$\sum\limits_i\frac{1}{\lambda_i}{(\frac{\partial\lambda_i}{\partial
D})}^2 +\frac{\partial^2\lambda_i}{\partial D^2}ln{\lambda_i}<0$;
iii) $D=mn\sqrt{\lambda_1\lambda_2\cdots\lambda_n}=
\frac{mn}{\sqrt{2}}\sqrt{{I_0^2-I_1}}$,
that is, $\prod\limits_i
f_i(u,v)=\frac{1}{2m}[(m-1)+m^2\sum\limits_{i\neq
j}f_i(u,v)f_j(u,v)]$. We call a mixed state $\rho$ given by
(\ref{rho}) $D$-computable if all the decompositions of $\rho$ on
pure states belonging to $\Psi$.

Due to the conditions i) and ii), for a $D$-computable state
$\rho$, calculating $E(\rho)$ is then reduced to the calculation
of the corresponding minimum of $D(\rho) = \mbox{min}\,
\sum_{a=1}^M p_a D(|\psi_a \rangle)$, which simplifies the
calculation if $D(|\psi_a \rangle)$ has a simpler expression than
$E(|\psi_a \rangle)$. The condition iii) guarantees that
$D$ is a quadratic form of the entries of the
matrix $A$ and can be expressed in the form of $D=|\langle
\psi|S\psi^*\rangle|$ in terms of a suitable matrix $S$, which
allows us to find an explicit analytical
expression of the entanglement of formation in a way similar to
the one used in \cite{HillWootters} and \cite{sjxg}. It generalizes
the results in \cite{sjxg,feili} where the case that
$AA^\dag$ has two non-zero eigenvalues is considered.

Let $S^{ipjq}$ be a symmetric $N^2\times N^2$ matrix whose
elements are all zero except for
$$\begin{array}{l}
S_{p+N(i-1),q+N(j-1)}=S_{q+N(j-1),p+N(i-1)}=1,\\
S_{q+N(i-1),p+N(j-1)}=S_{p+N(j-1),q+N(i-1)}=-1,
\end{array}$$
where $i,j,p,q=1,...,N$. Let $\Lambda_1^{ipjq}$,
$\Lambda_2^{ipjq}$, $\Lambda_3^{ipjq}$ and $\Lambda_4^{ipjq}$, in
decreasing order, be the eigenvalues of the rank four Hermitian
matrix $\sqrt{\sqrt{\rho}S^{ipjq}{\rho^\ast}S^{ipjq}\sqrt{\rho}}$.

{\sf [Theorem]}. For a $D$-computable state $\rho$, the minimum
decomposition of the generalized concurrence $D(\rho)$, i.e. the
average generalized concurrence of the pure states of the
decomposition, minimized over all decompositions of $\rho$, is
given by \be\label{drho}
\frac{mn}{4}[\sum_{i,j,p,q=1}^N(\Lambda_1^{ipjq}-\Lambda_2^{ipjq}-
\Lambda_3^{ipjq}-\Lambda_4^{ipjq})^2]^{\frac{1}{2}}. \ee

{\sf [Proof]}.
Let $r$ be the rank of $\rho$ and $|v_k\rangle$, $k=1,...,r$, be a
complete set of orthogonal eigenvectors corresponding to the
nonzero eigenvalues of $\rho$, such that $\langle v_k|v_k \rangle$
is equal to the $k$th eigenvalue. Other decomposition $\{
|w_k\rangle \}$ of $\rho$ can then be obtained through unitary
transformations:
\begin{equation}\label{w}
| w_k \rangle = \sum_{l=1}^r U_{kl}^* | v_l\rangle,
\end{equation}
where $U$ is a $t \times t$ unitary matrix, $t\geq r$. We have
$\langle w_k|S^{ipjq}{w}_l^*\rangle=(U \tau^{ipjq} U^T)_{kl}$,
where the matrix $\tau^{ipjq}$ is defined by
$\tau_{kl}^{ipjq}=\langle v_k|S^{ipjq}v_l^*\rangle$. As the matrix
$S^{ipjq}$ is symmetric, $\tau^{ipjq}$ is also symmetric and can
always be diagonalized by a unitary matrix $U$ such that $U
\tau^{ipjq} U^T=diag (\Lambda_1^{ipjq},...,\Lambda_r^{ipjq})$
\cite{HJ}. The diagonal elements $\Lambda_\alpha^{ipjq}$,
$\alpha=1,...,r$, in decreasing order, can always be made to be
real and non-negative. They are also the eigenvalues of the
Hermitian matrix $R \equiv
\sqrt{\sqrt{\rho}S^{ipjq}{\rho^\ast}S^{ipjq}\sqrt{\rho}}$. The
matrix $S^{ipjq}$ has $N^2-4$ rows and $N^2-4$ columns that are
identically zero. It can be seen that the corresponding rows and
columns of matrix $S^{ipjq}\rho^*S^{ipjq}$ are identically zero as
well. Thus the $N^2\times N^2$ Hermitian matrix
$S^{ipjq}\rho^*S^{ipjq}$ has a rank no greater than four. It
follows that the Hermitian matrix
$\sqrt{\sqrt{\rho}S^{ipjq}{\rho^\ast}S^{ipjq}\sqrt{\rho}}$ is
ranked at most four: at least
$\Lambda_5^{ipjq},\Lambda_6^{ipjq},...,\Lambda_r^{ipjq}$ are zero.

There always exits a decomposition consisting of states
$|w_k\rangle$, $k=1,\ldots ,r$, such that $\langle
w_k|S^{ipjq}{w}_l^*\rangle= \Lambda_k^{ipjq}\delta_{kl}$. Set
$|y_1\rangle = |w_1\rangle$, $|y_l\rangle = i|w_l\rangle$ for $l =
2 , ... , r$. Any decomposition can be written in terms of the
states $|y_k\rangle$ via the equation $| z_k \rangle =
\sum_{l=1}^r V_{kl}^* | y_l\rangle$, where $V$ is a $t \times r$
matrix whose $r$ columns are orthonormal vectors.

Denote $d_{ipjq}\equiv|\langle\psi|S^{ipjq}\psi^*\rangle|=
2|a_{ip}a_{jq}-a_{iq}a_{jp}|$.
The average $d_{ipjq}$ of a general decomposition is given by
\begin{equation}
\langle d_{ipjq}\rangle = \sum_k |(VYV^T)_{kk}|=\sum_k \Bigl|
\sum_l (V_{kl})^2 Y_{ll} \Bigr|,
\end{equation}
where $Y$ is the real diagonal matrix defined by $Y_{kl} = \langle
y_k|S^{ipjq}y_l^* \rangle$. Using the fact that $\sum_k
|(V_{kl})^2| = 1$, one gets
$$
\langle d_{ipjq}\rangle \ge | \sum_{kl} (V_{kl})^2 Y_{ll} | \ge
{\Lambda_1}^{ipjq}-{\Lambda_2}^{ipjq}-{\Lambda_3}^{ipjq}-{\Lambda_4}^{ipjq}.
$$
Since
$$
D(\vert\psi\rangle)=\frac{mn}{2}
\sqrt{\sum\limits_{i,j,p,q=1}^N|a_{ip}a_{jq}-a_{iq}a_{jp}|^2}
=\frac{mn}{4}\sqrt{\sum\limits_{i,j,p,q=1}^N d_{ipjq}^2},
$$
using the Cauchy-Schwarz inequality,
$$\sum\limits_{a=1}^M(\sum\limits_{i,j,p,q=1}^N
|\langle\psi_a|S^{ipjq}|\psi_a^*\rangle|^2)^{\frac{1}{2}}
\geq
[\sum\limits_{i,j,p,q=1}^N(\sum\limits_{a=1}^M|\langle\psi_a|S^{ipjq}|\psi_a^*\rangle|
)^2]^{\frac{1}{2}},
$$
we get
$$
\langle D(\rho)\rangle\geq\frac{mn}{4}(\sum_{i,j,p,q=1}^N \langle
d_{ipjq}\rangle^2)^{\frac{1}{2}}\,.
$$
Therefore the minimum decomposition
of the generalized concurrence $D(\rho)$ is given by
$$
\frac{mn}{4}[\sum_{i,j,p,q=1}^N(\Lambda_1^{ipjq}-\Lambda_2^{ipjq}-
\Lambda_3^{ipjq}-\Lambda_4^{ipjq})^2]^{\frac{1}{2}}.
$$
Due to convex relation between $E(\vert\psi\rangle)$ and $D(\vert\psi\rangle)$,
the entanglement of formation of $\rho$ is given by $E(D(\rho))$.
\hfill $\rule{2mm}{2mm}$

As a simple example we consider a class of pure states on $3\times 3$
(which is not the case in \cite{sjxg,feili})
with the matrix $A$ given by
\be\label{a} A=\left(\begin{array}{ccc}
a_{11}&a_{12}&a_{13}\\[3mm]
a_{21}&a_{22}&a_{23}\\[3mm]
a_{21}&a_{22}&a_{23}\\[3mm]
\end{array}\right).
\ee
The matrix $AA^\dag$ has two non-zero eigenvalues $\lambda_1$ and $\lambda_2$ satisfying
$$\lambda_1\lambda_2=2(|a_{11}a_{22}-a_{12}a_{21}|^2+|a_{11}a_{23}-a_{13}a_{21}|^2
+|a_{12}a_{23}-a_{13}a_{22}|^2).$$
It is directly verified that
all pure states given by (\ref{a}) belong to $\Psi$.

For all mixed states $\rho$ with decompositions on pure states (\ref{a}) (it is easily seen
that once $\rho$ has a decomposition with all the pure states given by (\ref{a}), then all
other kinds of decompositions belong to $\Psi$ too), from the theorem
the lower bound of the generalized concurrence $D(\rho)$ is given by
$$\sqrt{2}[(\Lambda_1^{1122}-\Lambda_2^{1122}-\Lambda_3^{1122}-\Lambda_4^{1122})^2
+(\Lambda_1^{1123}-\Lambda_2^{1123}-\Lambda_3^{1123}-\Lambda_4^{1123})^2
+(\Lambda_1^{1223}-\Lambda_2^{1223}-\Lambda_3^{1223}-\Lambda_4^{1223})^2]^{\frac{1}{2}}.
$$

We have studied the entanglement of formation for higher
dimensional quantum mixed states. It has been shown that under
certain conditions the entanglement of formation is a
monotonically increasing and convex function of a generalized
concurrence. For a class of ($D$-computable) arbitrary dimensional
mixed states an explicit lower bound of the entanglement of
formation is derived. The generalized concurrence defined in this
note is a generalization of the one in \cite{sjxg}. However, as we
imposed on the condition $\prod\limits_i
f_i(u,v)=\frac{1}{2m}[(m-1)+m^2\sum\limits_{i\neq
j}f_i(u,v)f_j(u,v)]$ in calculating the lower bound of the
entanglement of formation, the $D$-computable states here only
cover a part of the $d$-computable states in \cite{feili}. In
fact, in order to get the lower bound of the concurrence by using
the method above, it is not necessary for $D$ to be of the form
$\sum_{i,j,p,q=1}^N \vert a_{ip}a_{jq}-a_{iq}a_{jp}\vert^2$, but a
form of $\sum \vert F(a)\vert^2$ would be enough, where $F(a)$
stands for an arbitrary quadratic polynomial of the elements
$a_{ij}$.

We have assumed that $\lambda_i\neq 0$, $i=1,...n$,
in our theorem. For $n>1$, all the pure states in $\Psi$
are entangled. Therefore the $D$-computable states are not
separable in this case, as all possible decompositions,
including the minimum decomposition of the entanglement
of formation are in $\Psi$. One can check if these states
are bound entangled by simply checking whether their are PPT
(positive partial transposition) or not.

\end{document}